\documentclass[final]{siamltex}

\usepackage[pdfauthor={SJC & PHT}]{hyperref}
\usepackage[usenames, dvipsnames]{xcolor}
\hypersetup{colorlinks=true,linkcolor=MidnightBlue,citecolor=MidnightBlue}

\usepackage{graphicx}
\usepackage{amsmath, amssymb}

\usepackage{relsize}
\usepackage{pifont}
\usepackage{oubraces}
\usepackage{tikz}
\usepackage{listings}
\usepackage{booktabs}
\usepackage{xfrac}

\usepackage{tabularx}
\usepackage{array}
\usepackage{multirow}

\usepackage{mathrsfs}
\usepackage{microtype}

\usepackage{tikz}

\usepackage[numbers]{natbib}

\newcommand\eps{\epsilon}

\usepackage{pgfplotstable}

\usepackage{xparse}
\makeatletter
\NewDocumentCommand{\raisedminus}{m}{%
  \raisebox{0.05em}{$\m@th#1{-}$}%
}

\makeatother

\title{exponential asymptotics for the eigenvalues in the broken $\mathcal{PT}$-symmetric region}

\author{S. Jonathan Chapman\thanks{Oxford Centre for Industrial and
    Applied Mathematics, Mathematical Institute, Andrew Wiles
    Building, Woodstock Road, Oxford, Oxfordshire OX2 6GG, UK} \and Philippe H. Trinh\thanks{Department of Mathematical Sciences, University of Bath, Bath, Somerset BA7 7AY, UK.}}

\renewcommand*{\Re}{\operatorname{Re}}
\renewcommand*{\Im}{\operatorname{Im}}
\newcommand*{\Ai}{\operatorname{Ai}}

\newcommand*{\PT}{\mathcal{PT}}
\newcommand*{\q}{\varepsilon}
\newcommand{\e}{\mathrm{e}}

\newcommand{\Oh}{\mathcal{O}}

\newcommand*{\ep}{\epsilon}

\newcommand*{\ra}{\to}

\renewcommand*{\i}{\mathrm{i}}

\newcommand*{\de}{\operatorname{d\!}{}} 
\newcommand{\dd}[2]{\frac{\de#1}{\de#2}}

\newcommand*{\fI}{f_{\textrm{I}}}
\newcommand*{\fII}{f_{\textrm{II}}}
\newcommand*{\fIII}{f_{\textrm{III}}}

\newcommand*{\PI}{\Phi_{\textrm{I}}}

\newcommand*{\PIII}{\Phi_{\textrm{III}}}

\begin{document}

\maketitle

\begin{abstract}
Stemming from the seminal work of Bender \& Boettcher in 1998 (Phys. Rev. Lett. vol. 80 pp.~5243--5246), there has been great interest in the study of $\PT$-symmetric models of quantum mechanics, where the primary focus is with the study of non-Hermitian Hamiltonians that nevertheless produce countably infinite sets of real-valued eigenvalues. One of the fundamental models of such a system is governed by the Hamiltonian $H = \hat{p}^2 + x^2(\i x)^{\q}$. In their work, Bender \& Boettcher proposed a WKB methodology for the prediction of the discrete eigenvalues in the so-called unbroken region of $\q > 0$. However, the authors noted that this methodology fails to predict those `broken' eigenvalues for $\q < 0$. Here, we shall explain why the traditional WKB methodology fails, and we shall demonstrate how eigenvalues for all relevant values of $\q$ can be predicted using techniques in exponential asymptotics. These predictions provide excellent agreement to exact numerical results over nearly the entire range of values. Moreover, such techniques can be extended to a much wider range $\PT$-symmetric problems.
\end{abstract}



\pagestyle{myheadings}
\thispagestyle{plain}
\markboth{CHAPMAN AND TRINH}{EXPONENTIAL ASYMPTOTICS AND THE BROKEN $\mathcal{PT}$-SYMMETRIC REGION}

\section{Introduction} 

In classical or quantum mechanics, the equations governing the time-evolution of a system can be derived from a Hamiltonian, $H$. It is a standard axiom in quantum mechanics that $H$ must be Hermitian, and thus its eigenvalues real. This is in connection with the assumption that measurements of the system must correspond to eigenvalues of $H$, and hence the Hermitian property guarantees real-valued outcomes from the model.

The question, however, is whether the Hermitian property of $H$ is a necessity or whether there exist more fundamental restrictions on the Hamiltonian that would nevertheless result in real-valued eigenvalues. Largely beginning with the work of Bender \& Boettcher~\cite{bender1998real} in 1998 (though similar ideas proliferated the literature going back to Dyson~\cite{dyson1952divergence}), there has been a great deal of interest in studying so-called $\PT$-symmetric theories, which generalise the Hermitian property of Hamiltonians. Such theories posit that as a fundamental assumption, $H$ should be invariant under a parity transformation, $\mathcal{P}$, and time reversal, $\mathcal{T}$.

The canonical model of a non-Hermitian $\mathcal{PT}$-symmetric Hamiltonian is given by $H = \hat{p}^2 + x^2(\i x)^{\q}$ and corresponds to the time-independent Schr\"{o}dinger eigenvalue problem for $\psi = \psi(x)$, given by
\begin{equation} \label{eq:schro}
  -\dd{^2 \psi}{x^2} + x^2(\i x)^\q \psi = E \psi,
\end{equation}
where $E$ is the eigenvalue. The boundary conditions for the above problem require that $\psi \to 0$ exponentially rapidly as $|x| \to \infty$ in a pair of adjacent wedges in the complex $x$-plane [see later \eqref{eq:maineig}]. The $\PT$-symmetric nature of \eqref{eq:schro} is equivalent to the fact that the equation is invariant under the transformation $x\mapsto -x$ and $\i \mapsto -\i$.

\begin{figure} \centering
\includegraphics[width=1.0\textwidth]{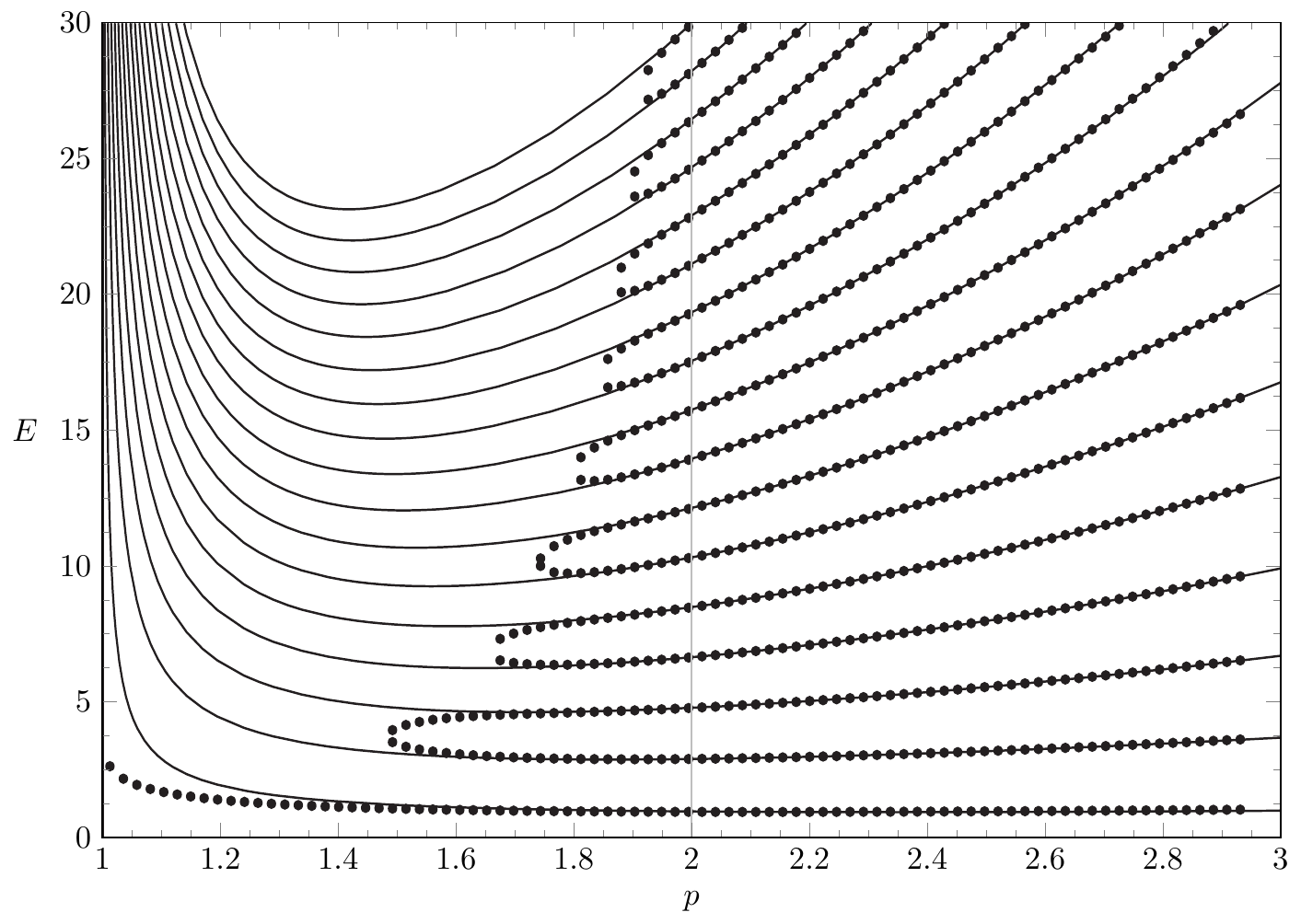}
\caption{Asymptotic eigenvalues of \eqref{eq:schro} given by \eqref{eq:benderEn} (lines) superimposed on the numerically calculated eigenvalues (dots). Note that here $p = 2 + \q$. The region of unbroken eigenvalues is for $p < 2$, where the `fingers' of the bifurcation curves first begin to close off. Note that the asymptotic approximation fails to predict the broken region. \label{fig:wkbeigs}}
\end{figure}

As explained in \emph{e.g.}~\cite{benderpt_book}, the remarkable significance of \eqref{eq:schro} is that for $\q > 0$, its eigenvalues, $E$, are real, despite the fact that the Hamiltonian is non-Hermitian. In particular, the following is now known about the spectrum of $H$, shown in Fig.~\ref{fig:wkbeigs}. 

\mbox{}\par
\begin{enumerate}
  \item For $\q > 0$, the eigenvalues, $E = E_n(\q)$, are real and form a discrete countably infinite set.  These eigenvalues match with the case of the harmonic oscillator, $E_n = 2n + 1$ at $\q = 0$. An asymptotic approximation of $E_n$ in the limit $n \to \infty$ was developed by Bender \& Boettcher~\cite{bender1998real}, and the reality of the spectrum was proved by Dorey \emph{et al.}~\cite{dorey2001spectral}. 
  \item It is also known, primarily through numerical solutions of the
    eigenvalue problem \eqref{eq:schro} that as $\q$ decreases below
    zero the eigenvalues move into the complex plane, forming complex-conjugate pairs.  The `fingers' in the bifurcation diagram begin to close off.
  \item There are further interesting behaviours in regards to the complex-valued eigenvalues and eigenfunctions for $\q < 0$. For example, there is an infinite-order exceptional point at $\q = -1$ where $|\Re E| \to \infty$ and $|\Im E| \to 0$ in the form of a logarithmic spiral. 
\end{enumerate}

\mbox{}\par


\noindent Our interest relates to the so-called broken region  $\q < 0$, and we highlight two open questions of significance. First, is there a simple explanation of why the spectrum must change at $\q = 0$ and why, for $\q < 0$, the `fingers' of the bifurcation diagram terminate? Second, is there an asymptotic approximation of the eigenvalues that remains valid in the broken region? 

\subsection{Key idea of why the eigenvalues terminate}

We find it convenient to set 
\begin{equation}
  p = 2 + \q,
\end{equation}
so that the exceptional point of $\q = 0$ corresponds to $p = 2$. 

In Bender \& Boettcher~\cite{bender1998real} it was shown that in the large eigenvalue limit, $|E| \to \infty$, the eigenvalues are given by [cf. their eqn (5)]
\begin{equation} \label{eq:benderEn}
  E = E_n = \left[ \frac{\sqrt{\pi}\left(n+\frac{1}{2}\right) \Gamma\left(\frac{3}{2} + \frac{1}{p}\right)}{\Gamma\left(1 + \frac{1}{p}\right)\sin\left(\frac{\pi}{p}\right)}\right]^{\frac{2p}{p+2}} \quad \text{for $n \to \infty$}.
\end{equation}
Bender \& Boettcher's derivation of \eqref{eq:benderEn} relies upon the development of distinct WKB approximations to \eqref{eq:schro}, each of which is valid as $|x|\to\infty$ in separate sectors of the complex plane. Matching the WKB solutions from one sector to the next involves a solvability condition that produces the eigenvalue conditions. If we introduce the small parameter
\begin{equation}
  \ep = E^{-\frac{p+2}{2p}},
\end{equation}
so that $\ep \to 0$ as $E\to\infty$, then in our notation, the solvability condition is
\begin{equation} \label{eq:big2}
  2\i\exp\left[\frac{2R(p)\cos\bigl(\frac{\pi}{p}\bigr)}{\ep}\right]\cos\left[\frac{2R(p)\sin\bigl(\frac{\pi}{p}\bigr)}{\ep}\right] = 0,
\end{equation}
where we have defined
\begin{equation} \label{eq:R}
  R(p) = \frac{\sqrt{\pi} \Gamma\left(1 + \frac{1}{p}\right)}{2\Gamma\left(\frac{3}{2} + \frac{1}{p}\right)}.
\end{equation}
Setting the argument of the cosine term in \eqref{eq:big2} to $(n + 1/2)\pi$ for $n\in\mathbb{Z}$ results in the Bender \& Boettcher asymptotic result of \eqref{eq:benderEn}. The WKB analysis that produces \eqref{eq:big2} mirrors the well-known semi-classical approach~\cite{muller2012introduction}, with the distinction of requiring integration of the differential equation in the complex plane. However, Bender \& Boettcher~\cite{bender1998real} note that for $p < 2$, the path of continuation along which the WKB solutions are typically matched must proceed through a branch cut in the complex plane; hence apparently condition \eqref{eq:big2} is no longer valid. We clarify this remark in \S\ref{sec:traditional}.

In this work, we shall demonstrate that for $p < 2$ (or $\q < 0)$, the eigenvalue condition \eqref{eq:big2} should instead include an additional term,
\begin{equation} \label{eq:big3}
   2\i\exp\left[\frac{2R(p)\cos\bigl(\frac{\pi}{p}\bigr)}{\ep}\right]\cos\left[\frac{2R(p)\sin\bigl(\frac{\pi}{p}\bigr)}{\ep}\right] - \frac{2\pi \i \ep^{p}}{2^{p + 2} \Gamma(-p)} = 0.
\end{equation}
When $p > 2$, then $R(p)\cos(\pi/p) > 0$ and the contribution from the second term of \eqref{eq:big3} is exponentially small in comparison to the first; thus \eqref{eq:benderEn} remain valid. However, for $1 < p < 2$ the dominance exchanges with $R(p)\cos(\pi/p) < 0$, and the second term now exponentially dominates. As $\ep \to 0$, there is no way to zero this second term, and hence in the large eigenvalue limit, there are no real eigenvalues that satisfy \eqref{eq:big3}. 

Consequently the new solvability condition \eqref{eq:big3} provides a formal argument of why the discrete family of eigenvalues must terminate for $p < 2$, with the `fingers' in the bifurcation diagram of Figure~\ref{fig:wkbeigs} closing firstly for the largest eigenvalues. Moreover, note that for finite values of $\ep$, it is still possible to satisfy \eqref{eq:big3}. The solution of this transcendental equation then allows for an asymptotic prediction of the eigenvalues. The agreement between asymptotically calculated eigenvalues and exact numerical eigenvalues is excellent, even at moderate values of the eigenvalue (see the later Figure~\ref{fig:fulleigsfig}).


The rest of this paper will be devoted to presenting the methodology that leads to the above results; in particular, we wish to emphasise that this methodology, which uses exponential asymptotics, is considerably general, and can likely be applied to the range of linear eigenvalue problems of interest to studies in $\PT$-symmetry. In fact, as we discuss in Sec.~\ref{sec:discussion}, similar methodologies are applied to much more difficult problems in nonlinear differential equations where standard WKB approaches are not valid.

\section{Mathematical formulation}

Recall our choice of $p = \q + 2$, and in anticipation of studying the
large eigenvalue limit of \eqref{eq:schro}, with $|E| \to \infty$, we
re-scale the independent variable in \eqref{eq:schro} by setting 
\begin{equation} \label{eq:xEscale}
  x = E^{1/p} z \qquad \text{and} \qquad \ep = E^{-\frac{p+2}{2p}},
\end{equation}
The $\mathcal{PT}$-symmetric eigenvalue problem for $\psi(x) = f(z)$ where $z\in\mathbb{C}$ is now given by
\begin{subequations} \label{eq:maineig}
\begin{gather}
- \eps^2 f''(z)   - (\i z)^p f(z) =  f(z), \label{eq:maineq} \\
\text{with $f \ra 0$ as $z \ra \infty \,\e^{\i\pi[-1/2 \mp 2/(p+2)]}$}, \label{eq:mainbc}
\end{gather}
\end{subequations}
where we are primarily interested in values of real values of $p \geq
1$. Above and in the remainder of this paper, primes ($'$) denote
differentiation with respect to $z$. 

\begin{figure}[t]\centering
\includegraphics[width=0.9\textwidth]{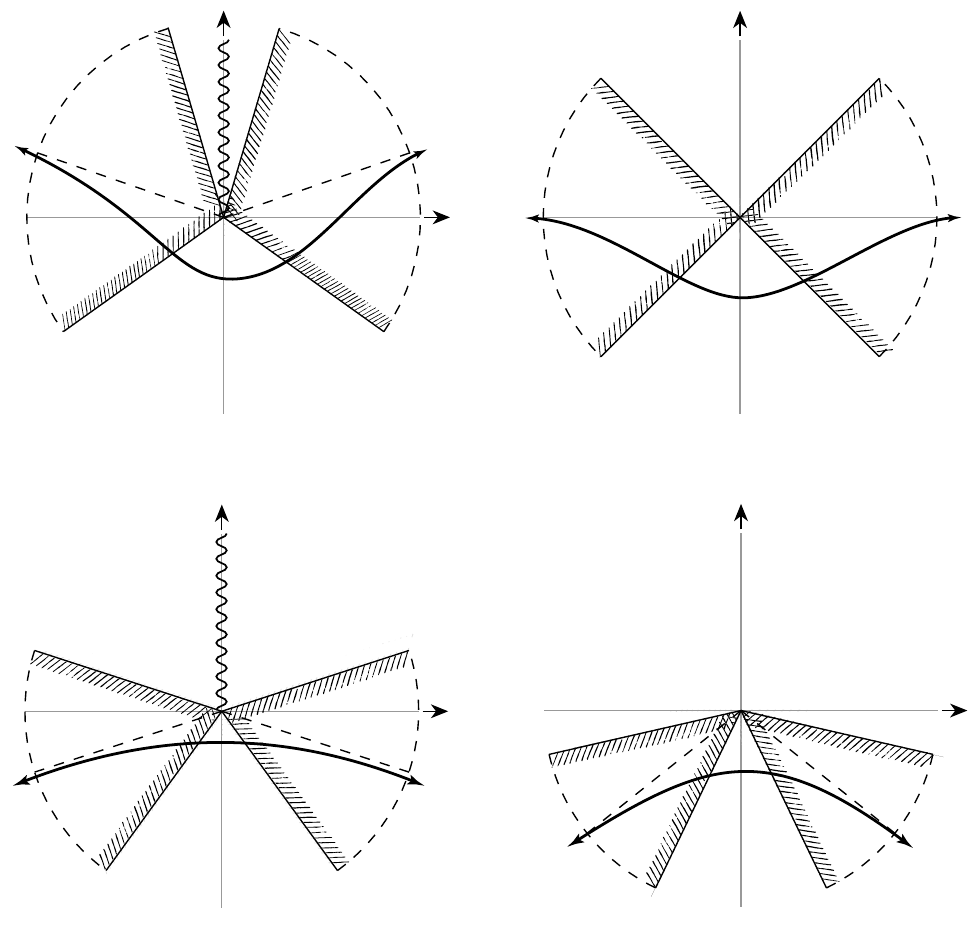}
\caption{Stokes wedges (hatched) for the $\mathcal{PT}$-symmetric problem \eqref{eq:maineig} shown in the $(\Re z, \, \Im z)$-plane for (a) $p = 1.3$; (b) $p = 2$; (c) $p = 2.9$; (d) $p = 5$. Note that if $p$ is not an integer, there is a branch point (wavy line) from $z = 0$. The thick curve indicates a continuation path defining the solution of \eqref{eq:maineq} which tends to the centre of the Stokes wedges (dashed) and given in \eqref{eq:thetac}. \label{fig:stokeswedge}}
\end{figure}

Let us explain the boundary conditions. The path on which we solve the above equation is shown in Fig.~\ref{fig:stokeswedge} and corresponds to the boundary conditions \eqref{eq:mainbc}. Note that for large $|z|$, the dominant balance of \eqref{eq:maineq} involves $\ep^2 f'' \sim -(\i z)^p f$. Consequently a WKB solution can be developed in the limit $|z| \to \infty$ and contains the exponential factor $\exp[\pm (\i z)^\alpha/(\alpha\epsilon)]$ where $\alpha = p/2+1$. When $\alpha$ is non-integral, consider the positive branch of $(\i z)^\alpha$ with the branch cut taken vertically upwards. With $z = |z|\e^{\i \theta}$ the exponential corresponding to this branch decays within two wedges (known as the \emph{Stokes wedges}) centred at angles
\begin{equation} \label{eq:thetac}
\theta_{\text{left/right}} = \pi \left(-\frac{1}{2} \mp \frac{2}{p+2}\right), 
\end{equation}
as shown in Fig.~\ref{fig:stokeswedge}. The size of each wedge is $2\pi/(p+2)$. The exponential argument that appears in the boundary condition \eqref{eq:mainbc} corresponds to the centre of the wedge, where the WKB solution is decaying most rapidly at infinity.

Note that a non-trivial solution to \eqref{eq:maineq} can be obtained by enforcing the condition that $f$ decays in any two non-adjacent Stokes sectors. The two particular sectors chosen in \eqref{eq:maineig} are such that the solution of the system is an analytic continuation of the case of the harmonic oscillator. For $p = 2$ there are two wedges of size $\pi/2$ and the solution is required to tend to zero as $|z| \to \infty$ along the real axis. For values of $p$ larger-than $p = 2$, the two wedges move into the lower-half plane, and coalesce along the negative imaginary axis when $p \to \infty$. For values of $p$ smaller than $p = 2$, the wedges move into the upper half-plane.

\section{Failure of the traditional WKB approach} \label{sec:traditional}

We first give an explanation of the traditional WKB approach, as applied in \emph{e.g.} Bender \& Boettcher~\cite{bender1998real}, and explain why this approach apparently fails for the case of the broken eigenvalue region $p < 2$. 

In the limit $\ep \to 0$, we approximate the solution using the WKB ansatz,
\begin{equation}\label{eq:WKB}
 f(z) \sim  e^{\i \phi(z)/\eps}\sum_{n=0}^\infty \eps^n A_n(z).
\end{equation}
Substitution into \eqref{eq:maineig} gives an eikonal equation for $\phi$ at leading order and an amplitude equation for $A_0$ at next order,
\begin{subequations}
\begin{align}
(\phi')^2 - (\i z)^p &= 1, \\
2\i \phi' A_0' + \i \phi'' A_0 &= 0.
\end{align}
\end{subequations}
These are solved to give
\begin{subequations} \label{eq:WKB_phiA0}
\begin{align}
  \phi(z) &= \pm \int^z [1  +(\i t)^p]^{1/2}\, \de{t}, \label{eq:phi_WKB} \\
  A_0(z) &= \frac{\text{const.}}{(\phi')^{1/2}} = \frac{\text{const.}}{[1 + (\i z)^p]^{1/4}}. \label{eq:nonharmA0}
\end{align}
\end{subequations}

From \eqref{eq:WKB_phiA0}, the WKB approximation fails at those turning points where 
\begin{equation} \label{eq:tpeqn}
  1+(\i z)^p=0.
\end{equation}
Of the turning points in \eqref{eq:tpeqn} only two lie in the appropriate sectors of the complex plane and are relevant for the analysis: 
\begin{equation} \label{eq:zAB}
  z_A = -\i\e^{-\i\pi/p} \qquad \text{and} \qquad 
  z_B = -\i \e^{\i\pi/p}.
\end{equation}
The fact that the other turning points are not involved at leading-order will become clear in the exponential asymptotics methodology presented later. 



We may write down a composite WKB approximation. From \eqref{eq:phi_WKB}, we choose the positive sign of $\phi$ and define 
\begin{equation}
\begin{gathered}
\PI(z) = \int_{z_A}^z [1  +(\i t)^p]^{1/2}\, \de{t}, \qquad
\PIII(z) = \int_{z_B}^z [1  +(\i t)^p]^{1/2}\, \de{t}, \\
\Phi(z) = \int_a^z [1  +(\i t)^p]^{1/2}\, \de{t},
\end{gathered}
\end{equation}
where $a$ can be taken to be any point where the integral is defined (\emph{e.g.} $a = 0$). The principal branches are chosen throughout. Then we form the following composite solution:
\begin{equation}\label{star}
 f(z) \sim 
\begin{cases}
\displaystyle
\biggl[ \frac{a_1}{(\phi')^{1/2}}\biggr] \e^{\i\PI/\ep}
 + \biggl[\frac{b_1}{(\phi')^{1/2}}\biggr] \e^{-\i\PI/\ep} & \text{for $z$ in Region I}, \\[0.4cm]
 \displaystyle
 \biggl[\frac{a_2}{(\phi')^{1/2}}\biggr] \e^{-\i\Phi/\ep} + 
\biggl[\frac{b_2}{(\phi')^{1/2}}\biggr] \e^{\i\Phi/\ep} & \text{for $z$ in Region II},\\[0.4cm]
\displaystyle
 \biggl[\frac{a_3}{(\phi')^{1/2}}\biggr] \e^{-\i\PIII/\ep} +
 \biggl[\frac{b_3}{(\phi')^{1/2}}\biggr] \e^{\i\PIII/\ep} & \text{for $z$ in Region III}.
\end{cases}
\end{equation}

\begin{figure}\centering
\includegraphics{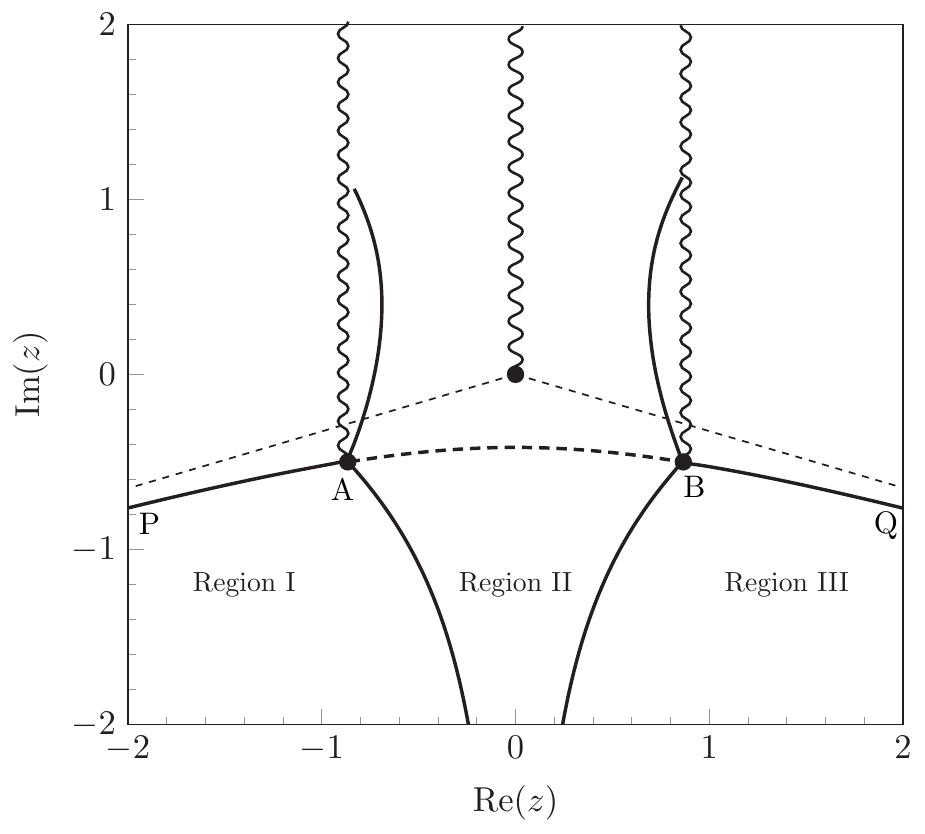}
\caption{Stokes line configuration for $p = 3$. From each of the two turning points at A and B, three equal-phase lines $\Im(\i\PI) = 0 = \Im(\i\PIII)$ are drawn (thick solid). There is a curve (thick dashed) where $\Re(\i\PI) = 0 = \Re(\i\PIII)$ along which the traditional WKB matching procedure of \S\ref{sec:traditional} is performed. Branch cuts are shown as wavy lines. The thin dashed line corresponds to the centre of the Stokes wedge. \label{fig:stokesp3}}
\end{figure}

The three regions are shown in Figure~\ref{fig:stokesp3} for the case of $p = 3$. From each of the two turning points, marked A and B, there are three solid lines, indicating curves where $\Im(\i\PI) = 0$ and $\Im(\i\PIII) = 0$, respectively. Along PA, $(\i\PI)$ is purely real and negative and similarly along BQ, $(\i\PIII)$ is real and negative. Thus, examining the composite solution \eqref{star}, in order for $f$ to decay as $|z| \to \infty$ along PA and BQ, we require $b_1 = a_3 = 0$. 

This leaves four unknown constants. A local analysis near the turning points, $z = z_A$, lets us relate solutions in Region I with II and hence $\{a_1, \, b_2\}$ to $\{a_2, \, b_2\}$. Similarly solutions in Region II and III are matched from a local analysis near $z = z_B$, and this relates $\{a_2, \, b_2\}$ to $\{a_3, b_3\}$. The details of this matching procedure are given in Appendix~\ref{sec:harm_constants}. This local asymptotic analysis re-scales the independent variable, $z$, such that locally near $z = z_{B}$ say, the solution, $f$, behaves as an Airy function to leading order. The asymptotic behaviour of the (real-valued) Airy function at positive and negative infinity then allows the solutions in Regions II and III to be matched. Crucially, this matching occurs along the segment AB show in Figure~\ref{fig:stokesp3} where $\Re(\i\PI) = \Re(\i\PIII)$. 

The result is a pair of homogeneous linear equations 
\begin{subequations}
\begin{align}
a_2 \e^{\i\Phi(B)/\eps } - \i b_2 \e^{-\i\Phi(B)/\eps} &= 0, \label{eq1} \\
 a_2 \e^{\i\Phi(A)/\eps } + \i b_2 \e^{-\i\Phi(A)/\eps } &= 0, \label{eq2}
\end{align}
\end{subequations}
for the two unknowns $a_2$ and $b_2$, and where we have used the shorthand $\Phi(A) = \Phi(z_A)$ and $\Phi(B) = \Phi(z_B)$. Thus, in order for a nonzero
solution to exist we need
\begin{equation}
\left| \begin{array}{cc}
\e^{\i\Phi(B)/\eps } & - \i  \e^{-\i\Phi(B)/\eps }\\
\e^{\i\Phi(A)/\eps } & \i  \e^{-\i\Phi(A)/\eps }
\end{array} \right| = \i\e^{\i[\Phi(B) - \Phi(A)]/\eps} + 
\i\e^{-\i [\Phi(B) - \Phi(A)]/\eps} = 0.
\end{equation}
This gives the eigenvalue condition 
\begin{equation} \label{eq:select0orig}
\i \e^{2\i\Phi(A)/\eps} + \i \e^{2\i\Phi(B)/\eps} = 0 \Longrightarrow \e^{2\i[\Phi(B) - \Phi(A)]/\eps   + \i \pi} = 1,  
\end{equation}
or in terms of $\phi$ in \eqref{eq:phi_WKB}, 
\begin{equation}
\frac{2}{\eps} \Bigl[ \phi(B) - \phi(A) \Bigr] = 
\frac{2}{\eps} \int_{z_A}^{z_B} (1+(\i t)^p)^{1/2}\, \de{t} = (2n + 1) \pi \qquad \text{for $n \in\mathbb{Z}$}.  
\end{equation}
Evaluating the integral explicitly, and remembering that $\eps = E^{1/2 + 1/p}$ from \eqref{eq:xEscale}, returns the Bender \& Boettcher~\cite{bender1998real} result of \eqref{eq:benderEn}. 

As we had noted earlier in the introduction, the eigenvalues given by this formula are illustrated in Figure~\ref{fig:wkbeigs}, along with a numerical calculation of the exact eigenvalues. We see that, where the eigenvalues exist, the approximation \eqref{eq:benderEn} is good. However, it fails to find that there are only a finite number of eigenvalues for some values of $p$.

\begin{figure}\centering
\includegraphics{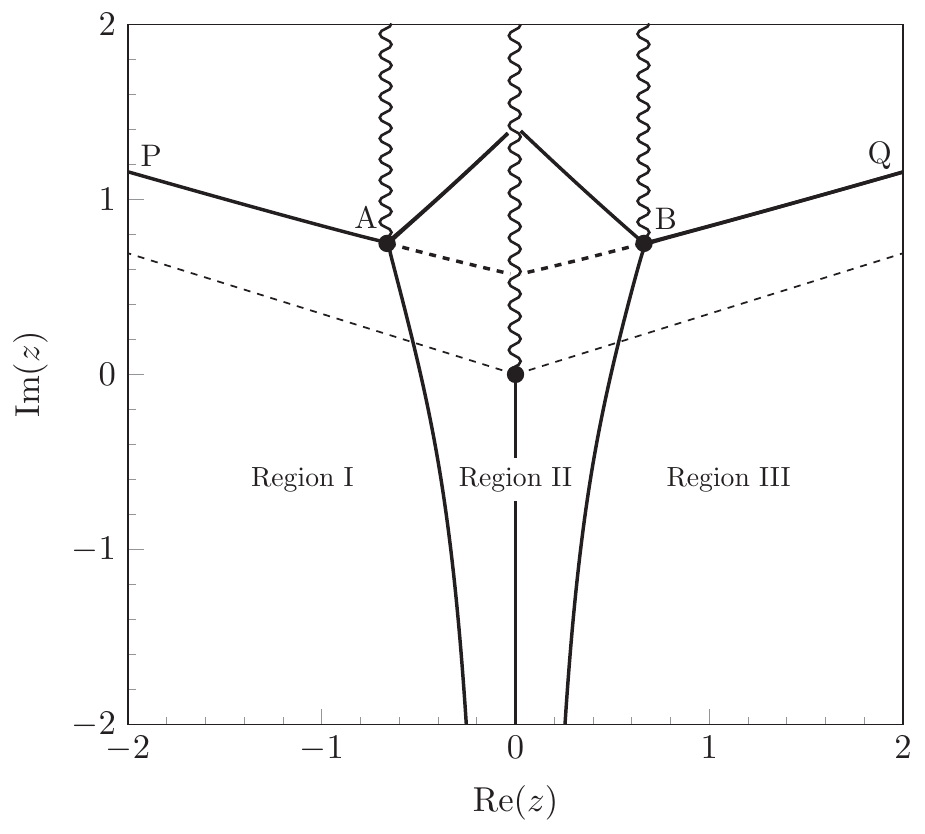}
\caption{Stokes line configuration for $p = 1.3$. From each of the two turning points at A and B, three equal-phase lines $\Im(\i\PI) = 0 = \Im(\i\PIII)$ are drawn (thick solid). There is a curve (thick dashed) where $\Re(\i\PI) = 0 = \Re(\i\PIII)$ along which the traditional WKB matching procedure of \S\ref{sec:traditional} is performed. Branch cuts are shown as wavy lines. The thin dashed line corresponds to the centre of the Stokes wedge. \label{fig:stokesp1p3}}
\end{figure}

The WKB approximation fails due to the fact that when $p$ is not an integer there is a branch point at the origin, with a branch cut up the positive imaginary axis, say. For $p>2$ the path between $z_A$ and $z_B$ on which $\phi$ is real misses this branch cut (it passes below the origin, as shown in Figure~\ref{fig:stokesp3}, and the WKB analysis gives a good approximation to the eigenvalues. However, when $p<2$ the path between $z_A$ and $z_B$ on which $\phi$ is real passes through the branch cut, as shown in Figure~\ref{fig:stokesp1p3}. Thus the turning points which have been connected via the WKB analysis (and thus the points at infinity at which the boundary conditions are imposed) lie on different Riemann sheets of the solution. 

In principle, this traditional WKB approach can be modified by taking
the first WKB approximation, analytically continuing it around the
branch point at $z = 0$, and then matching with the second WKB
approximation. However, to do so would already require a knowledge of
Stokes switchings in the WKB approximation, so that it is 
more straightforward to simply keep track of the Stokes lines on the
original Riemann sheet.

\section{An alternative approach using Stokes lines} 
\label{sec:harm_exp}

In the traditional WKB approach, solutions are developed on different subregions of the plane, and then matched together. In this case, the eigenvalues $\ep = \ep_n$ emerge as a result of solvability conditions on the constants of integration. However, we saw that this approach does not allow a prediction of the eigenvalues in the region $p < 2$. In this section, we present an approach that uses exponential asymptotics. In this approach, the discrete spectrum \eqref{eq:benderEn} arises in connection with the Stokes phenomenon, and we shall find an additional contribution that causes the bifurcation curves to close. 

We return to the WKB approximation in \eqref{eq:WKB} and this time, we choose a single exponential argument from \eqref{eq:phi_WKB}, with
\begin{subequations} \label{eq:phiA0}
\begin{align}
\phi(z) &=  \int_{a}^z [1 +(\i t)^p]^{1/2}\, \de{t} \label{eq:phi} \\
A_0(z) &= \frac{1}{[\phi'(z)]^{1/2}}, \label{eq:A0_again}
\end{align}
\end{subequations}
%
%
where $a$ can be freely chosen. For simplicity, the constant  in \eqref{eq:nonharmA0} is taken to be unity. Thus we consider the leading-order approximation 
\begin{equation} \label{eq:star2}
  f(z) \sim \frac{1}{[1 +(\i z)^p]^{1/4}} \exp\left[\frac{1}{\ep} \int_a^z [1 +(\i t)^p]^{1/2}\, \de{t}\right].
\end{equation}

Above in \eqref{eq:phi} $\phi$ has been defined with the positive square-root branch in mind (say with both branch cuts from $z = z_{A}, z_B$ taken upwards). We may verify that as $|z|\to \infty$ along the centre of the Stokes wedges, the WKB ansatz \eqref{eq:star2} decays with $f \to 0$. As we know, the WKB solution is singular at the two turning points, and this necessitated the matched procedure presented in the last section. 

However, it is still possible to travel from one wedge to the other without passing through the two singularities. If $z$ remains below the curve PABQ in Figure~\ref{fig:stokesp3}, then there is no apparent indication that \eqref{eq:star2} should fail to approximate the true solution, $f$, over the desired path and indeed, \eqref{eq:star2} seems to be valid whatever the value of $\eps$. One arrives at the erroneous conclusion that there is no restriction on $\ep$ and a continuous spectrum results.

The apparent paradox is resolved by taking into account the extra terms that are switched-on due to the Stokes Phenomenon. In what follows, we shall see that as \eqref{eq:star2} is analytically continued from $-\infty$ to $+\infty$, two subdominant exponentials are switched-on---one due to a Stokes line from $z = -1$ and another due to a Stokes line from $z = 1$. The subsequent restriction of these exponentials in order to satisfy the boundary conditions is what produces the discrete set of eigenvalues.







\subsection{Exponential asymptotics} \label{sec:expasym}

We require exponential asymptotics to determine those exponentially small terms and the conditions for their switching, and this can be done using a procedure of optimal truncation and Stokes-line smoothing~\cite{chapman_1998_exponential_asymptotics}. For convenience, let us write $f(z) = \e^{\i\phi/\ep} A(z)$ so that the differential equation \eqref{eq:maineq} is now 
\begin{equation} \label{eq:Aeq}
  \eps A'' + 2\i \phi' A' + \i \phi'' A = 0.
\end{equation}
We then proceed further by expanding fully
\begin{equation}\label{eq:Aseries}
 A(z) \sim  \sum_{n=0}^\infty \eps^n A_n(z).
\end{equation}
At $\Oh(1)$, $A_0$ is given by \eqref{eq:A0_again} as we know. At $\Oh(\eps^n)$ in \eqref{eq:Aeq}, the equation for $A_n$ is given by
\begin{equation} \label{An} 
 A_{n-1}'' + 2 \i  \phi' A_{n}' + \i A_{n} \phi'' = 0.
\end{equation}


Notice that $A_0$ in \eqref{eq:A0_again} has singularities at the
turning points, $1 + (\i z)^p = 0$, as well as at $z = 0$ if $p$ is
not an integer. However, solving for $A_n$ involves differentiation of the previous order, $A_{n-1}$. Thus generically, the power of each singularity must grow at each subsequent order and consequently $A_n$ diverges in the form of a factorial over a power as $n \to \infty$. Such divergence is typical of singular perturbation problems \citep{chapman_1998_exponential_asymptotics,dingle_book}. We assume the late terms diverge in the form of 
\begin{equation} \label{eq:ansatz}
  A_n(z) \sim \frac{B(z) \Gamma(n+\gamma)}{[\chi(z)]^{n+\gamma}} \quad \text{as $n\to\infty$}
\end{equation}
with $\chi = 0$ at the singularities of the early terms. Note that a
separate factorial-over-power ansatz of the form \eqref{eq:ansatz} is
required for each singularity in the late terms, but only those singularities that are associated with active Stokes lines need to be accounted for. Substituting \eqref{eq:ansatz} into \eqref{An} gives, for the first two orders in the limit $n \to \infty$,
\begin{subequations}
\begin{gather}
-2 \i \chi' \phi' + (\chi')^2 = 0, \label{eq:ueq} \\
2 \i B' \phi' + \i B \phi'' - 2 \chi' B' - \chi B'' = 0. \label{eq:Beq}
\end{gather}
\end{subequations}

The first equation in \eqref{eq:ueq} yields $\chi(z)$, a quantity known as the \emph{singulant}~\cite{dingle_book}:
\begin{equation} \label{eq:u}
  \chi(z) = 2 \i \bigl[\phi(z) - \phi(z_*)\bigr],
\end{equation}
and by assumption, $\chi = 0$ at those singularities $z = z_*$ that cause $A_n$ to diverge. By linearity of the asymptotic procedure no new singularities are introduced apart from those in the early terms. Hence $z_*$ must correspond either to the turning points given by \eqref{eq:tpeqn} or, for non-integral powers $p$, the branch point at $z = 0$. 

At next order \eqref{eq:Beq} is solved, giving
\begin{equation} \label{eq:harmB}
B(z) = \frac{\Lambda}{(\phi')^{1/2}}, \qquad \Lambda = \mbox{constant}.  
\end{equation}
Hence the late terms in the WKB expansion of $f$ are
\begin{equation}
\eps^n A_n \e^{\i \phi/\eps} \sim  \eps^n \left[\frac{\Lambda}{(\phi')^{1/2}} 
\frac{\Gamma(n+\gamma)}{[2\i (\phi-\phi(z_*))]^{n+\gamma}}\right] \e^{\i \phi/\eps} \quad \mbox{ as
}n \ra \infty
.\label{late} 
\end{equation}
As noted by \emph{e.g.} Dingle~\cite{dingle_book}, this divergence of
the expansion is associated with subdominant  exponentials that are
switched-on via the Stokes Phenomenon.  

\subsection{Inner-matching procedure}

As it will be shown from analysis of the Stokes lines in \S\ref{sec:expswitch}, only three of the singularities are important for specification of the eigenvalues: the two former points introduced as $z_A$ and $z_B$ in \eqref{eq:zAB} and the branch point $z = 0$. There are thus three factorial/power ansatzes of the form \eqref{late} to consider, and values of $\gamma$ and $\Lambda$ must be determined for each case. 

First, the value of $\gamma$ can be found by ensuring that the order of the singularity in $A_n$ \eqref{eq:ansatz} is consistent with that of $A_0$ in \eqref{eq:A0_again}. Next, the value of $\Lambda$ can be found by developing an inner solution valid near the singularities and matching with the WKB solution. The matching procedure detailed in Appendix~\ref{sec:appendixmatch} shows how these are derived. In summary, they are given by
\begin{subequations}\label{eq:gen_gammaLam}
\begin{alignat}{4}
z_A &= -\i\e^{-\pi\i/p}, &\qquad \chi_A &= 2\i[\phi - \phi(z_A)], &\qquad \gamma &= 0, &\qquad \Lambda &= \frac{1}{2\pi}, \\
z_B &= -\i\e^{\pi\i/p}, &\qquad \chi_B &= 2\i[\phi - \phi(z_B)], &\qquad \gamma &= 0, &\qquad \Lambda &= \frac{1}{2\pi}, \\
z &= 0, &\qquad \chi_0 &= 2\i[\phi - \phi(0)], &\qquad \gamma &= -p, &\qquad \Lambda &= -\frac{1}{2^{p+2}\Gamma(-p)}. \label{eq:gamLam3}
\end{alignat}
\end{subequations}
We have now completely determined the behaviour of the late terms, of which the relevant divergence is driven by a sum of three factorial/power ansatzes of the form \eqref{eq:ansatz} or alternatively \eqref{late}. The values of $\chi(z)$, $B(z)$, $\gamma$, and $\Lambda$ are given by \eqref{eq:u}, \eqref{eq:harmB}, and \eqref{eq:gen_gammaLam}. 



\subsection{Exponential switchings} \label{sec:expswitch}

The connection between the late terms \eqref{late} and the exponentials switched-on from the Stokes Phenomenon can be understood based on a procedure of optimal truncation and Stokes-line smoothing \cite{berry1990waves,chapman_1998_exponential_asymptotics,dingle_book}. Briefly, the idea is as follows. First, the solution to \eqref{eq:Aeq} is expressed as a truncated expansion plus a remainder term, 
\begin{equation}
  A(z) \sim  \sum_{n=0}^{N-1} \eps^n A_n(z) + R_N(z),
\end{equation}
so that the equation for the remainder satisfies
\begin{equation} \label{eq:Rneq}
  \ep R_N'' + 2\i \phi' R_N' + \i \phi'' R_N \sim -\ep^N A_{N-1}''.
\end{equation}
In the limit $\eps \to 0$, the optimal truncation point, $N \to \infty$. Thus the behaviour of the remainder is predicated by the behaviour of the late terms, thus establishing a connection between the exponentials switched on and the divergence of the series. If the series is optimally truncated, then $R_N$ is exponentially small. Moreover, the exponentially small remainder is switched-on when the solution is analytically continued across critical curves in the complex plane known as \emph{Stokes lines} in a process known as the \emph{Stokes Phenomenon}. Analysis of \eqref{eq:Rneq} indicates two crucial facts that have been demonstrated by \emph{e.g.} \cite{berry1990waves,chapman_1998_exponential_asymptotics,dingle_book}. 

First, there are Stokes lines where successive late terms of \eqref{eq:Aseries} have the same phase, i.e. those points $z\in\mathbb{C}$ where $u$ is positive and real. Thus, 
\begin{equation} \label{eq:stokesline}
\Im[\chi(z)] = 0 \quad \text{and} \quad \Re[\chi(z)] \geq 0,  
\end{equation}
where $u$ is given by \eqref{eq:u}.

Second, the exponentials that are switched-on are given by
\begin{equation} \label{eq:Aexp}
  A_\mathrm{exp} \sim \frac{2\pi \i}{\eps^\gamma} B(z)\e^{-\chi(z)/\eps} = \frac{2\pi \i}{\eps^\gamma} \frac{\Lambda}{(\phi')^{1/2}} \e^{-\chi(z)/\eps},
\end{equation}
and a derivation of this result is given in Appendix~\ref{sec:switch}.

Thus, our goal is to consider the paths of continuation shown in Figure~\ref{fig:stokeswedge} and determine which Stokes lines from which singularities are crossed. As the base asymptotic series \eqref{eq:Aseries} is analytically continued across each Stokes line, it must switch-on the exponentially small term \eqref{eq:Aexp} where $\chi$, $B$, and $\gamma$ correspond to the singularity from which the Stokes line originates. 

As shown in Figure~\ref{fig:stokesp3}, for general non-integral values of $p$, there are three sets of relevant Stokes lines. Moving through the three Regions I, II, and III thus switches on the three contributions,
\begin{equation} \label{eq:switch1}
  \left[\i \e^{-\chi_A(z)/\ep} + \i \e^{-\chi_B(z)/\ep}\right] - \frac{2\pi \i \ep^p}{2^{p+2} \Gamma(-p)} \e^{-\chi_0(z)/\ep},
\end{equation}
where we have substituted values of $\gamma$ and $\Lambda$ from \eqref{eq:gen_gammaLam}. Notice that the exponential arguments are proportional to $\e^{-2\i\phi}$. Following the discussion of \eqref{star}, we note that $\Re(-2\i\phi)$ is real and negative as $|z|$ tends to infinity along the (Anti-Stokes) line marked BQ in Figures~\ref{fig:stokesp3} and \ref{fig:stokesp1p3}. Thus, although the terms in \eqref{eq:switch1} exponentially small at the point they are switched on, they become exponentially large as $|z| \to \infty$ in Region III. For the solution to decay it is necessary for \eqref{eq:switch1} to be zero. Using $\chi$ from \eqref{eq:gen_gammaLam}, we have
\begin{equation}\label{fulleigs}
 \i e^{2 \i \phi(z_A)/\eps} + \i e^{2 \i \phi(z_B)/\eps} - \frac{2 \pi \i
\eps^p}{2^{p+2} 
\Gamma(-p)}e^{2 \i \phi(0)/\ep} = 0.
\end{equation}


Note that by direct integration, 
\begin{equation}
  \phi(z_A) - \phi(0) = \int_0^{z_A} [1 + (\i x)^p]^{1/2} \, \de{x} = \left[-\sin(\pi/p) - \i \cos(\pi/p)\right] R(p),
\end{equation}
where $R$ is defined in \eqref{eq:R}, and moreover $\phi(z_B) = -\overline{\phi(z_A)}$. Combination of the two first terms in \eqref{fulleigs} then yields the same equation as first stated in \eqref{eq:big3}, or 
\begin{equation} \label{eq:big3_again}
  2\i \exp\left[\frac{2R(p)\cos\bigl(\frac{\pi}{p}\bigr)}{\ep}\right]\cos\left[\frac{2R(p)\sin\bigl(\frac{\pi}{p}\bigr)}{\ep}\right] - \frac{2\pi \i \ep^{p}}{2^{p + 2} \Gamma(-p)} = 0.
\end{equation}
The eigenvalue condition \eqref{eq:big3} is the main result. 

Notice that for $p>2$, $R(p)\cos(\pi/p) > 0$ and the first term exponentially dominants the second. Setting the cosine term to zero hence yields the Bender \& Boettcher~\cite{bender1998real} result of \eqref{eq:benderEn}. In other words, this is equivalent to the condition that the exponential switched-on across the Stokes line from $z_A$ is switched off due to the Stokes line from $z_B$. The classical quantisation condition has been re-derived.

\begin{figure} \centering
\includegraphics[width=1.0\textwidth]{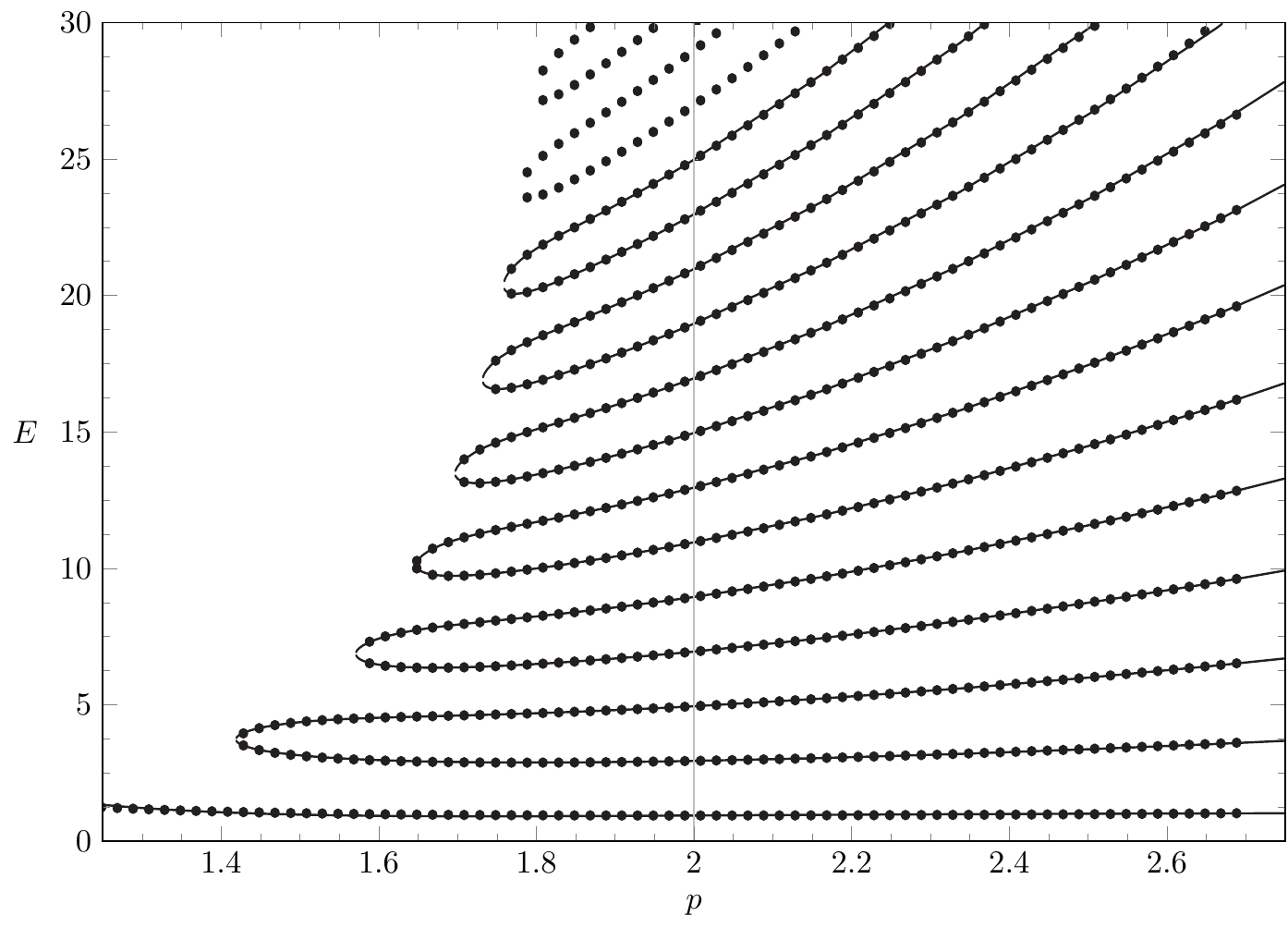}
\caption{Asymptotic eigenvalues given by (\ref{fulleigs}) (lines) overlaid with the numerically calculated eigenvalues (dots) \label{fig:fulleigsfig}}
\end{figure}

However, for for $p<2$, $R(p)\cos(\pi/p) < 0$ and the exponential dominance switches, with the second term in \eqref{eq:big3_again} now dominant. In the limit $\eps \ra 0$, there are no solutions; thus there are only a finite number of real eigenvalues for $p<2$. The roots of equation (\ref{fulleigs}) are shown in Figure~\ref{fig:fulleigsfig}, along with the numerical calculation of the eigenvalues. The agreement is very good.

\subsection*{Extension to complex eigenvalues}

\begin{figure}\centering
\includegraphics[width=1.0\textwidth]{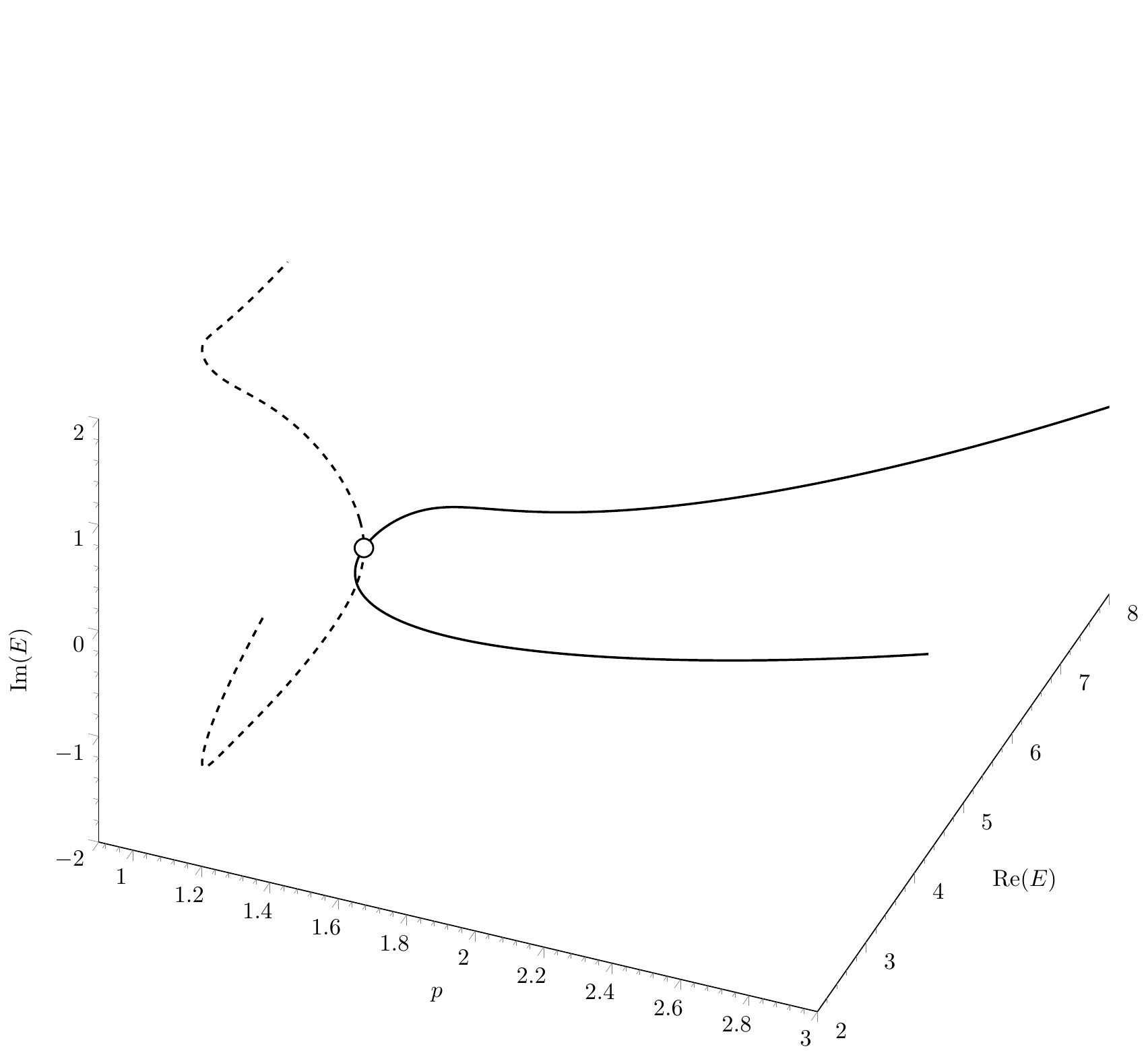}
\caption{Extension of the eigenvalue curves into the complex plane using \eqref{eq:big3_again}. \label{fig:complexbranch}}
\end{figure}

As the branches of the bifurcation diagram coalesce (as in Fig.~\ref{fig:fulleigsfig}), the real-valued eigenvalues merge in pairs; thereafter, for smaller values of $p$, there are two associated complex-conjugate eigenvalues. For the case of the two merging braches going through $E = 1$ and $E =3$ and $p = 2$, this produces the curves shown in Fig.~\ref{fig:complexbranch}.

In a recent paper, Bender \emph{et al.}~\cite{bender2017behavior} have presented numerical and asymptotic results  for the eigenvalues of \eqref{eq:schro} (with $p = 2 + \q$) near the special values of $p = \{1, -1, -2\}$. In fact, the exponential asymptotics we have presented \S\ref{sec:expasym} can be used to reproduce many of their results in a more unified way. 

\begin{figure} \centering
\includegraphics{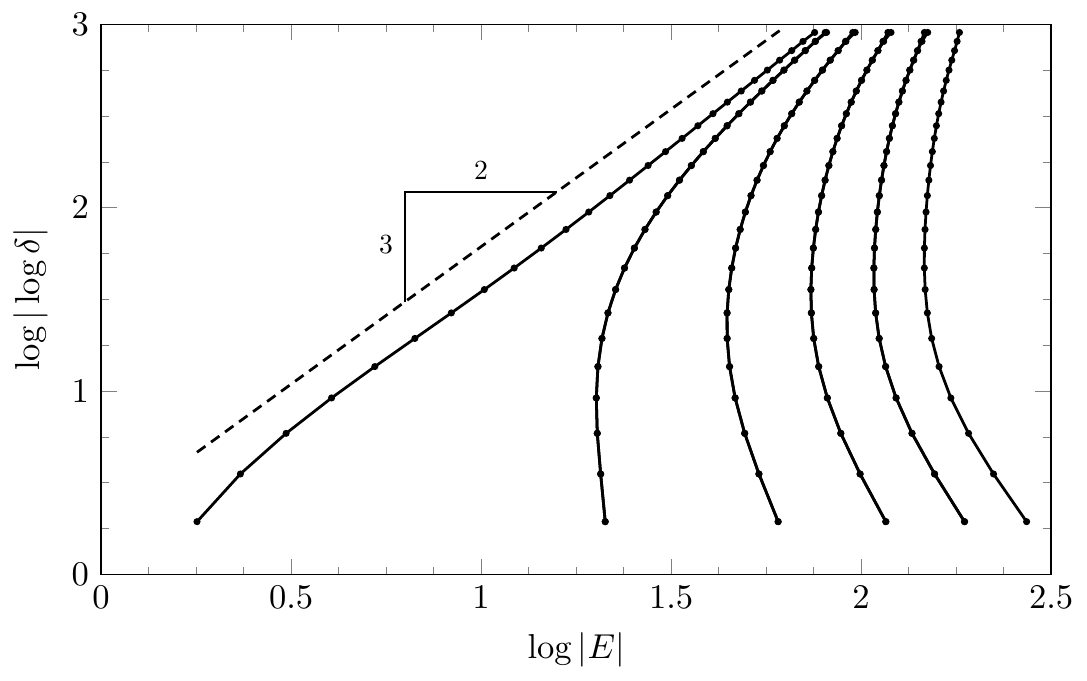}
\caption{As $\delta \to 0$ where $p = 1 + \delta$, we expect that $\delta \sim (8/\pi)E^{3/2}\exp[-(4/3)E^{3/2}]$. Hence on the graph, the numerically calculated eigenvalues (solid markers) should tend to the dashed line of $\log|\log \delta| \sim (3/2)\log|E| + \log(-4/3)$. The eigenvalues shown correspond to the lowest six branches. \label{fig:loglogdelt}}
\end{figure}

\begin{figure} \centering
\includegraphics{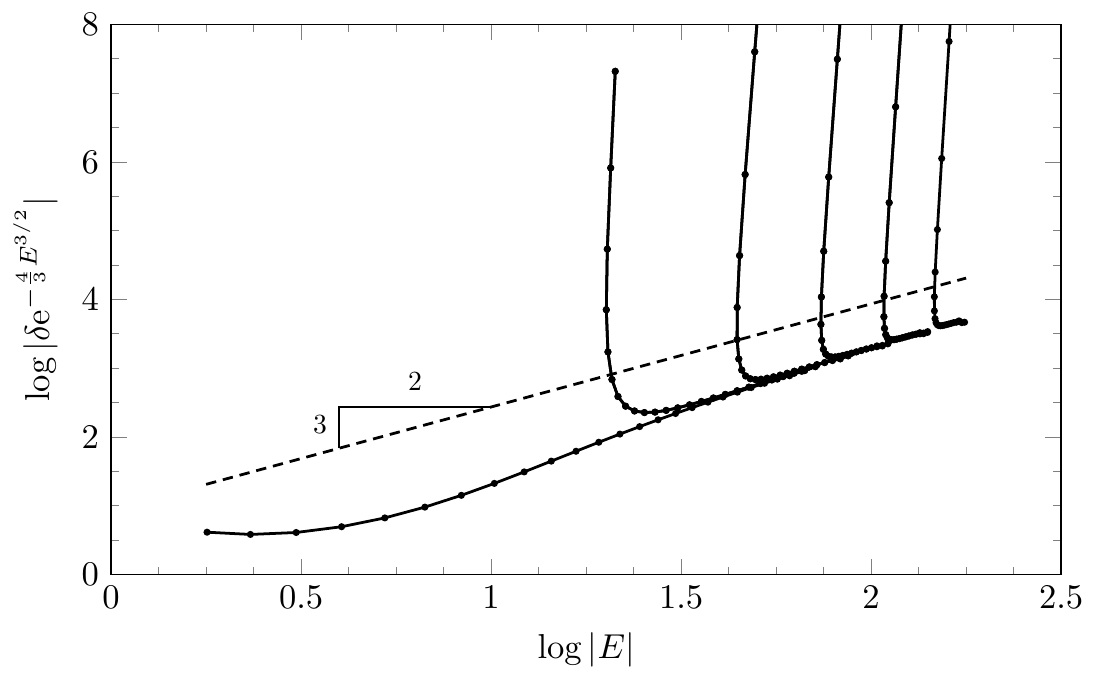}
\caption{As $\delta \to 0$ where $p = 1 + \delta$, we expect that $\delta \exp(4E^{3/2}/3) \sim (8/\pi)E^{3/2}$. Hence on the graph, the numerically calculated eigenvalues (solid markers) should tend to the dashed line of $3/2 \log |E| + \log(8/\pi)$. \label{fig:logdeltexp}}
\end{figure}

By way of example, let us investigate the limit of $p\to1^+$, where $|\Im E| \to 0$ and $|\Re E| \to \infty$. In this limit, the two turning points, labeled A and B, in Fig.~\ref{fig:stokesp1p3} coalesce at $z = \i$ and consequently, the width of Region II, bounded by the two Stokes lines, must shrink. Although turning points A and B coalesce, the previous Stokes-line contributions we have derived for points A, B, and C remains valid and it may be verified that the inner-region asymptotics of Appendix~\ref{sec:appendixmatch} remain unaltered in this limit. Thus we may let $p = 1 + \delta$ for $\delta \to 0^+$ directly in the eigenvalue condition \eqref{eq:big3_again}. Then $R(p) \sim 2/3$ and $1/\Gamma(-p) \sim \delta$, and
\begin{equation}
  2\exp\left(-\frac{4}{3\ep}\right) \sim \frac{\pi}{4} \ep \delta.
\end{equation}
Since $\ep \sim E^{-3/2}$ by \eqref{eq:xEscale}, we have in terms of $E$,
\begin{equation} \label{eq:delta}
  \delta \sim \frac{8E^{3/2}}{\pi}\exp\left(-\frac{4}{3} E^{3/2}\right).
\end{equation}
Fig.~\ref{fig:loglogdelt} provides verification that the exponential scaling predicted in \eqref{eq:delta} agrees with the numerical calculation of the eigenvalues in the limit $p \to 1^+$ and $\delta \to 0$. Similarly Fig.~\ref{fig:logdeltexp} provides verification that the lower-order algebraic dependence of $E^{3/2}$ is as expected. Note that in providing the exact pre-factor and algebraic scaling, our \eqref{eq:delta} provides a more accurate prediction of the eigenvalues than asymptotic formula in eqn (12) of \cite{bender2017behavior}. 

\section{Application to other $\PT$-Ssymmetric problems}

The exponential asymptotic techniques we have presented provide a more powerful and general framework than the traditional WKB analysis of \eqref{sec:traditional}. Thus, this idea of locating singularities in the asymptotic expansions, and examining the switching-on of exponentials as Stokes lines are crossed can be used to study the broken and unbroken eigenvalues of wider range of $\PT$-symmetric problems than the case of \eqref{eq:maineig}. 



\begin{figure}[htb] \centering
\includegraphics{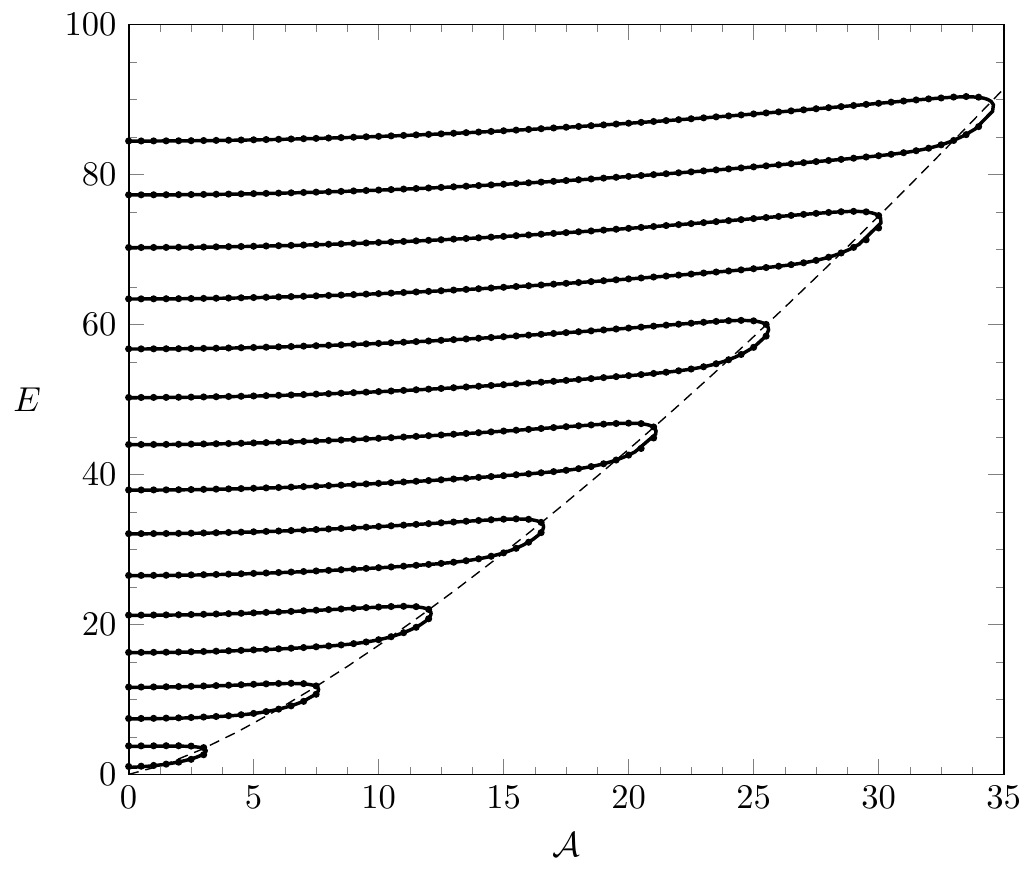}
\caption{Eigenvalues, $E$, of the quartic oscillator \eqref{eq:quartic} as a function of the parameter $\mathcal{A}$. The eigenvalues are identical for $\mathcal{A} < 0$. The $E\to\infty$ asymptotic solutions (solid) are generated from \eqref{eq:select4} and are nearly visually indistinguishable from the numerical solutions (circles). It is expected that the bifurcation curves close-off at approximately $E \sim (\mathcal{A}/a^*)^{4/3}$ where $a^* \approx 1.18384$ (dashed). \label{fig:berry_eigs}}
\end{figure}

As another illustrative example, let us consider the Hamiltonian $H = \hat{p}^2 - x^4 - \i\mathcal{A}x$ that corresponds to the Schr\"{o}dinger equation
\begin{equation}
  -\dd{^2 \psi}{x^2} + (x^4 + \i \mathcal{A} x) \psi = E\psi,
\end{equation}
where $\mathcal{A}$ is a real parameter. The eigenvalues are shown in Fig.\ref{fig:berry_eigs} and they illustrate the familiar unbroken ($\mathcal{A}$ small) and broken ($\mathcal{A}$ large) regions we have observed in our previous example. Interestingly, the prediction of the eigenvalues in both regions were performed by Bender \emph{et al.}~\cite{bender2001complex} using traditional WKB techniques. Here, we shall show how the same analysis can be done through a straightforward application of the theory we have already developed. 

We shall seek to describe the bahaviour in the limit $|E| \to \infty$. Under the re-scaling $x = E^{1/4}z$, we have
\begin{equation} \label{eq:quartic}
-\ep^2 f''(z) + (z^4 + \i a z)\psi(z) = \psi(z).  
\end{equation}
where $a = \mathcal{A} E^{-3/4}$ and $\ep = E^{-3/4}$. We set $f(z) = \e^{\i\phi/\ep} A(z)$ and find that (taking the positive branch)
\begin{equation} \label{eq:phiberry}
  \phi = \int_b^z \left[ 1 - (t^4 + \i a t)\right]^{1/2} \, \de{t},
\end{equation}
where $b$ is an arbitrary point of integration. 

\begin{figure} \centering
\includegraphics[width=0.7\textwidth]{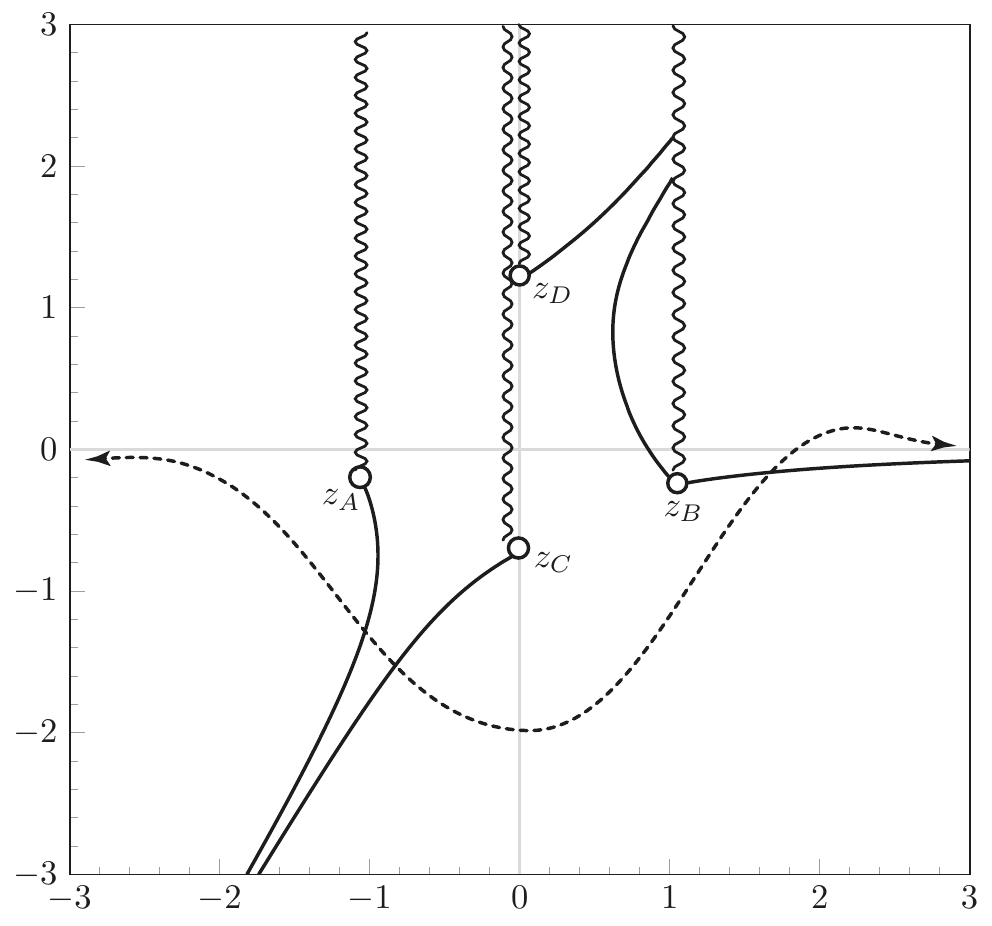}
\caption{Stokes lines for the quartic oscillator \eqref{eq:quartic} at $a = 1$. As the path of continuation (dashed) crosses Stokes lines (dashed) from turning points $z_A$, $z_C$, and $z_B$, the exponential switching of \eqref{eq:berry} occurs. Branch cuts are shown with the wavy line. \label{fig:berry}}
\end{figure}

The equation for $A(z)$ is identical to that of \eqref{eq:Aeq}. Thus, written in terms of $\phi$, the development of the late terms \eqref{eq:ansatz} and Stokes switching relation in \eqref{eq:Aexp} are identical. This time, the only singularities of the leading-order problem $A_0 \sim \text{const.}/(\phi')^{1/2}$ are those corresponding to turning points, where
\begin{equation}
  z^4 + \i a z = 1. 
\end{equation}
The four turning points are shown for the typical value of $a = 1$ in Figure~\ref{fig:berry}. For all real values of $a$, there are two roots along the imaginary axis, say $z_C$ and $z_D$, and two roots located at constant $\Im z$, say $z_A$ and $z_C$. The Stokes lines, where $\Im \chi = 0$ and $\Re\chi \geq 0$, computed from \eqref{eq:u} are plotted solid in the figure. The path of continuation considered by \cite{bender2001complex} is shown dashed. Thus the WKB solution is analytically continued across the three Stokes lines from $z_A$, $z_C$, and $z_B$, the switching is given by an expression analogous to \eqref{fulleigs}, with 
\begin{equation} \label{eq:berry}
  \i \e^{2\i\phi(z_A)/\ep} + \i \e^{2\i\phi(z_B)/\ep} + \i \e^{2\i\phi(z_C)/\ep} = 0.
\end{equation}
It is convenient to choose the point of integration in \eqref{eq:phiberry} as $b = z_C$, and consequently $\phi(z_B) = -\overline{\phi(z_A)}$. Dividing by the third exponential, we now have 
\begin{equation} \label{eq:select4}
  2 \e^{2V(a)/\ep} \cos\left[\frac{2U(a)}{\ep}\right] + 1 = 0,
\end{equation}
where we have defined
\begin{equation} \label{eq:UV}
  U(a) + \i V(a) = -\phi(z_A) = -\int_{z_C}^{z_A} \left[ 1 - (t^4 + \i a t)\right]^{1/2} \, \de{t},
\end{equation}
so that the notation is consistent to that of eqn (13) in \cite{bender2001complex}. From numerical integration, it can be verified that the function $V(a)$ in \eqref{eq:UV} begins at $V \approx 0.87402$ and decreases monotonically as $a$ increases, passing through $V = 0$ at $a = a^* \approx 1.18384$. 

Thus for $a < a^*$, the first term of \eqref{eq:select4} is exponentially large and the eigenvalues are predicted by those values of $\ep$ where $2U(a)/\ep \sim (2\mathbb{Z} + 1)/2$. This produces the countably infinite set seen in Fig.~\ref{fig:berry_eigs}. However, when $a > a^*$, the second term in \eqref{eq:select4} is now dominant and the bifurcation curves are expected to close-off. The fit between asymptotic and numerical results are shown in Fig.~\ref{fig:berry_eigs} and we see that even through the entire range of $E$-values, the agreement is nearly visually indistinguishable. Thus our results in this section have duplicated those of Bender \emph{et al.}~\cite{bender2001complex} in their study of the quartic oscillator \eqref{eq:quartic}, but here we have used the exponential asymptotic framework of Sec.~\ref{sec:harm_exp} instead of the traditional WKB methodology in Sec.~\ref{sec:traditional}. 

There is an important distinction between the main $\PT$-symmetric problem of this paper, given in \eqref{eq:maineig} corresponding to the Hamiltonian $H_1 = \hat{p}^2 - (\i x)^p$, and the sub-problem of this section, given in \eqref{eq:quartic}, with Hamiltonian $H_2 = \hat{p}^2 + x^4 + \i \mathcal{A} x$. The case of $H_1$ is unexpectedly challenging on account of the fact that the closing-off of eigenvalues is due to a singularity of a different (non-turning-point) type than those responsible for generating the main discrete set of eigenvalues. This necessitated the more powerful exponential asymptotic techniques we have proposed, and was perhaps the reason why the prediction of the eigenvalues of $H_1$ remained unresolved, in contrast to those of $H_2$. 

\section{Conclusion} \label{sec:discussion}

In this paper, we have shown how the eigenvalues of the Bender \& Boettcher~\cite{bender1998real} problem can be predicted in the broken region of $\q < 0$ (or $p < 2)$. Previous asymptotic analyses have relied on a traditional WKB framework of matching between turning points, and we have shown that this approach is inadequate. Instead, we have proposed a methodology that use exponential asymptotics to derive the eigenvalues in both broken and unbroken regions. Thus, as the relevant parameter decreases through the critical value of $\q = 0$, a previously subdominant contribution switches on and dominance exchanges. The match between the numerical and asymptotic values is excellent over the entire range of parameters. 

More generally, we have shown that the large eigenvalue asymptotics of linear eigenvalue problems may be understood in terms of Stokes lines in the complex plane, and that this interpretation provides a means of calculating the eigenvalues. A subdominant exponential is turned on across Stokes lines. This exponential grows at infinity, so that its coefficient must eventually be zero, which gives the eigenvalue condition. This may be thought of in terms of the contributions from two Stokes lines being exactly out of phase, and so destructively interfering. In the classical case (\emph{e.g.} corresponding to the harmonic oscillator) the eigenvalues form a series for which the corresponding eigenfunctions have an integer number of oscillations between the two Stokes lines, mirroring the picture on the real axis between the turning points, but in terms exponentially small ``beyond all orders''. However, for the case of general Hamiltonians, their associated singularity and Stokes-line structures may yield additional contributions that change the previous selection mechanism of eigenvalues. The Bender \& Boettcher~\cite{bender1998real} is one such case where such an additional contribution causes the eigenvalues to transition from unbroken (countably infinite) to broken (finitely many). It is expected that many other problems in $\PT$-symmetric quantum mechanics can be studied using this framework.

\begin{appendix}
\section{Determination of constants} \label{sec:harm_constants}

We first explain how \eqref{eq1} is derived by matching about the turning point near $z = z_B$, although much of the presentation applies to an arbitrary turning point at $z = z_*$. We-scale in the inner region with 
\begin{equation} \label{eq:tpsub}
  z = z_* + K\ep^{2/3} y,
\end{equation}
where $\i p (\i z_*)^{p-1} K^3 = -1$. The branch of $K$ is chosen so that $\phi$ in the later expression \eqref{eq:matchtp_phiphipu} corresponds to a decaying exponential, $\e^{\i \phi/\ep}$, as $y \to \infty$. Under this scaling, 
\begin{gather}
 \phi \sim \phi(z_*) + \i \ep (2/3) y^{3/2}, \\
  (\phi')^{1/2} \sim P \ep^{1/6} y^{1/4},
\end{gather}
where $P$ is a constant that depends on $K$, but whose exact value we do not require.

Under the substitution \eqref{eq:tpsub}, the main equation \eqref{eq:maineq} now yields the standard Airy equation
\begin{equation} \label{eq:newairy}
  \dd{^2 f}{y^2} - y f = 0.
\end{equation}
The solution of \eqref{eq:newairy} that matches with the requisite decay condition at infinity is $f = C \Ai(y)$ for a constant $C$. From \cite{abram}, we may write the full expansion of the Airy function as
\begin{equation} \label{eq:airyout}
 f = C \Ai(y) \sim  
 \frac{C\e^{-\frac{2}{3}y^{3/2}}}{2\sqrt{\pi}\, y^{1/4}},
\end{equation}
valid as $y \ra \infty$. This solution is then matched with the inner limit of the WKB solution $f = \fIII$ in \eqref{star}. In this limit, 
\begin{equation} \label{eq:fIII_inner}
  \fIII \sim \left[\frac{b_3}{P\ep^{1/6} y^{1/4} }\right] \e^{-\frac{2}{3} y^{3/2}}.
\end{equation}
Matching \eqref{eq:airyout} and \eqref{eq:fIII_inner} gives $C/(2\sqrt{\pi}) = b_3/(P \ep^{1/6})$. 

Next, we turn to the matching of the inner solution with the solution of Region II in \eqref{star}. Firstly, from \cite{abram}, we have as $y \to -\infty$,
\begin{equation} \label{eq:finner_outleft}
  C\Ai(y) \sim \frac{C\e^{\i\pi/4}}{\sqrt{\pi}(-y)^{1/4}} \left[ \e^{-\i \frac{2}{3}(-y)^{3/2}} - \i \e^{\i \frac{2}{3}(-y)^{3/2}}\right].
\end{equation}
This should be matched with the inner limit of $f = \fII$ in \eqref{star} as $z \to z_B$ and in fact, this is the only step in the above presentation that requires the use of $z_B$ instead of a generic $z_*$ for the turning point (since the definition of the solution in region III shifts the exponential argument to be zero at $z_B$). We have
\begin{equation} \label{eq:fII_in}
\fII \sim \left[\frac{a_2}{P \ep^{1/6} y^{1/4}}\right] \e^{-\i \Phi(B)/\ep} \e^{-\i \frac{2}{3} (-y)^{3/2}} 
+ \left[\frac{b_2}{P \ep^{1/6} y^{1/4}}\right] \e^{\i \Phi(B)/\ep} \e^{\i \frac{2}{3} (-y)^{3/2}},  
\end{equation}
and combining \eqref{eq:finner_outleft} with \eqref{eq:fII_in} gives the pair of equations 
\begin{subequations}
\begin{align}
  \left[\frac{C\e^{\i\pi/4}}{\sqrt{\pi}(-y)^{1/4}}\right] &= \left[\frac{1}{P \ep^{1/6} y^{1/4}} \right] a_2 \e^{-\i \Phi(B)/\ep}, \\
  \left[\frac{C\e^{\i\pi/4}}{\sqrt{\pi}(-y)^{1/4}}\right] (-\i) &= 
  \left[\frac{1}{P \ep^{1/6} y^{1/4}} \right] b_2 \e^{\i \Phi(B)/\ep},
\end{align}
\end{subequations}
for which $C$ can be eliminated to give 
\begin{equation} \label{eq1_pre}
  a_2 \e^{-\i \Phi(B)} - \i b_2 \e^{\i \Phi(B)} = 0.
\end{equation}
which is the desired \eqref{eq1}. Similarly, through a local analysis at $z=z_A$ we find
\begin{equation} \label{eq2_pre}
 a_2 \e^{\i\Phi(A)/\eps } + \i b_2 \e^{-\i\Phi(A)/\eps } = 0, 
\end{equation}
which is the desired \eqref{eq2}. In fact, the analysis near $z = z_A$ results in replacing the argument of $\Phi$ in \eqref{eq1_pre} with the turning point $z_A$ and morever swapping $a_2 \mapsto b_2$ and $b_2 \mapsto a_2$, which is a result of our definition of $\fI$ in \eqref{star}.

\section{Inner matching procedure for general $p$} \label{sec:appendixmatch}

The exponential asymptotics procedure requires deriving the form of the factorial/power ansatz of the late terms. From \eqref{eq:ansatz}, \eqref{eq:u}, \eqref{eq:harmB} we have
\begin{equation} \label{eq:epnAn_again}
  \ep^n A_n \sim \ep^n \frac{\Lambda \Gamma(n + \gamma)}{(\phi')^{1/2} [\chi(z)]^{n+\gamma}},
\end{equation}
where $\chi(z_*) = 0$ at the respective singularities given by $z = z_*$. The two numerical constants $\gamma$ and $\Lambda$ are determined by taking the limit of $z \to z_*$ and matching with the inner solution.

 \subsection{Derivation of $\gamma$ and $\Lambda$ for the turning points} \label{sec:matchtp}

It is somewhat easier to see the necessary re-scalings by examining the inner-region first. We let $z_*$ be a turning point and re-scale in the inner region with 
\begin{equation} \label{eq:xtpsub}
  z = z_* + K\ep^{2/3} y,
\end{equation}
where $\i p (\i z_*)^{p-1} K^3 = -1$. The branch of $K$ is chosen so that $\phi$ in the later expression \eqref{eq:matchtp_phiphipu} corresponds to a decaying exponential, $\e^{\i \phi/\ep}$, as $y \to \infty$. Under the substitution \eqref{eq:xtpsub}, the main equation \eqref{eq:maineq} now yields the standard Airy equation
\begin{equation} \label{eq:matchtp_airy}
  \dd{^2 f}{y^2} - y f = 0.
\end{equation}
The solution of \eqref{eq:matchtp_airy} that matches with the requisite decay condition at infinity is $f = C \Ai(y)$ for a constant $C$. From \cite{abram}, we may write the full expansion of the Airy function as
\begin{equation}
 f = C \Ai(y) \sim  
 \frac{C\e^{-\frac{2}{3}y^{3/2}}}{2\sqrt{\pi}\, y^{1/4}}\sum_{n=0}^\infty \frac{(-1)^n
\Gamma(3n+1/2)}{54^{n} \, n! \, \Gamma(n+1/2) (\frac{2}{3}y^{3/2})^n},\label{eq:inner1_edit}
\end{equation}
valid as $y \ra \infty$. 

Now with the outer limit of the inner solution, we return to the outer expansion and take the inner limit. First, using the scalings \eqref{eq:xtpsub} in \eqref{eq:phi} and \eqref{eq:u}, we have
\begin{equation} \label{eq:matchtp_phiphipu}
\begin{gathered}
  \phi \sim \phi(z_*) + \i \ep (2/3) y^{3/2}, \\
  (\phi')^{1/2} \sim P \ep^{1/6} y^{1/4}, \\
u \sim -(4/3)\ep y^{3/2},
\end{gathered}
\end{equation}
%
and where $P$ is an $\Oh(1)$ constant that we do not need to specify. Next, from \eqref{eq:phiA0}, the leading-order outer solution, $A_0$, written in the inner coordinates, satisfies
\begin{equation} \label{eq:matchtp_A0}
A_0 \sim \frac{1}{P \ep^{1/6} y^{1/4}},
\end{equation}
while substituting of the scalings \eqref{eq:matchtp_phiphipu} into the general late-orders expression \eqref{eq:epnAn_again} gives 
\begin{equation} \label{eq:matchtp_epAn}
\eps^n A_n \sim \frac{\Lambda \ep^{-\gamma}}{P \ep^{1/6} y^{1/4}} \frac{
  \Gamma(n+\gamma)}{(-\frac{4}{3}y^{3/2})^{n+\gamma}},
\end{equation}
For the power of the singularity in \eqref{eq:matchtp_A0} to match \eqref{eq:matchtp_epAn} at $n=0$, we require $\gamma =0$.

To determine $\Lambda$ we match with the outer solution with the inner solution \eqref{eq:inner1_edit} in the vicinity of the turning point. First, notice that the behaviour in \eqref{eq:inner1_edit} must match in the limit $y\to\infty$ with the leading-order outer WKB solution given by $A_0\e^{\i\phi/\ep}$. With $\phi$ from \eqref{eq:matchtp_phiphipu} and $A_0$ from \eqref{eq:matchtp_A0}, we have that 
\begin{equation} \label{eq:matchtp_C}
  C = \left[\frac{2\sqrt{\pi}}{P \ep^{1/6}}\right] \e^{\i \phi(z_*)/\ep}.
\end{equation}
We now match the $n$th term of the outer-to-inner limit with the $n$th term of the inner-to-outer limit. This gives
\begin{equation}
\frac{C\e^{-\frac{2}{3}y^{3/2}}}{2\sqrt{\pi}\, y^{1/4}}\left[\frac{(-1)^n
\Gamma(3n+1/2)}{54^{n} \, n! \, \Gamma(n+1/2) (\frac{2}{3}y^{3/2})^n}\right] \sim \frac{\Lambda}{P \ep^{1/6} y^{1/4}} \frac{
  \Gamma(n)}{(-\frac{4}{3}y^{3/2})^{n}},  
\end{equation}
Substitution of $C$ from \eqref{eq:matchtp_C} into the above yields the much more compact expression,
\begin{equation}
  \Lambda \sim \frac{1}{27^n} \frac{\Gamma(3n + \frac{1}{2})}{\Gamma(n)\Gamma(n_+1)\Gamma(n+\frac{1}{2})} = \frac{1}{2\pi}.
\end{equation}
Notice that had we chosen the constant of integration in $A_0$ in \eqref{eq:A0_again} to be a general value of $a_0$ (instead of $a_0 = 1$), then $\Lambda = a_0/(2\pi)$ above.

\subsection{Derivation of $\gamma$ and $\Lambda$ for the singularity $z = 0$} \label{sec:match0}
In order to determine the values of $\gamma$ and $\Lambda$ that correspond to the divergence due to $z = 0$, we apply a similar procedure to that of Sec.~\ref{sec:match0}. Here, the main difference is that the full solution in the inner region does not exist in terms of special functions.

In the limit $x \ra 0$, we have that $u \sim 2 \i x$, and thus from \eqref{eq:nonharmA0}, the leading-order prefactor behaves as 
\[ A_0 \sim 1 - \frac{(\i x)^p}{4} + \cdots.\]
\[ A_n \sim \frac{\Lambda
  \Gamma(n+\gamma)}{(2ix)^{n+\gamma}}.\] 
The order of the singularity is correct when $n=0$ only if $\gamma=-p$.
To determine $\Lambda$ we need to  match with an inner region in the
vicinity of $x=0$.
The inner scaling is $A = 1 + \eps^p g$,
$x = -i \eps y$, giving
\begin{equation}
\dd{^2 g}{y^2} + 2 \dd{g}{y} + \frac{p y^{p-1}}{2} = 0  
\end{equation}
at leading order, with the condition that $g \sim -y^p/4$ as
$y \ra \infty$.
We set $g = -y^p h/4$ to give
\begin{equation}
\biggl[\frac{p y}{2}\biggr] + \biggl[\frac{p(1-p)}{4}\biggr]h 
- \biggl[\frac{p y }{2}\biggr]h
- \biggl[\frac{p y} {2}\biggr]\dd{h}{y} - \biggl[\frac{y^2}{2}\biggr]\dd{h}{y} - \biggl[\frac{y^2}{4}\biggr] \dd{^2 h}{y^2} = 0,  
\end{equation}
with $h \ra 1$ as $y \ra \infty$. To match with the outer WKB solution expand as $y \ra  \infty $ as 
\begin{equation}
h(y) = \sum_{n=0}^\infty \frac{h_n}{y^n},  
\end{equation}
giving after simplification,
\begin{equation}
h_0 = 1, \qquad 
h_{n+1} = \left(\frac{n-p}{2}\right)h_n \qquad \text{for $n \geq 1$},  
\end{equation}
with solution $h_n = \Gamma(n-p)/[2^n \Gamma(-p)]$. Hence as $y \to \infty$, the outer limit of the leading-order inner solution is
\begin{equation} \label{eq:gen0_in2out}
A \sim 1 - \frac{\eps^p y^p}{4} \sum_{n=0}^\infty 
\frac{\Gamma(n-p)}{2^n \Gamma(-p) y^n}.  
\end{equation}

The above outer limit of the inner solution must be matched with the inner limit of the outer solution, $A \sim \sum \ep^n A_n$, whose $n$th term is 
\begin{equation} \label{eq:gen0_out2in}
\eps^n A_n \sim \frac{\eps^n\Lambda \Gamma(n-p)}{(2 \i x)^{n-p}}
  \sim \frac{\eps^p\Lambda \Gamma(n-p)}{(2 y)^{n-p}}.   
 \end{equation} 
Thus by the Van Dyke matching rule~\cite{vandyke_book}, matching \eqref{eq:gen0_in2out} with \eqref{eq:gen0_out2in} gives
\begin{equation}
\Lambda  = -\frac{1}{2^{p+2}} \lim_{n\ra \infty}
\frac{h_n 2^n}{\Gamma(n-p)} =  -\frac{1}{2^{p+2}\Gamma(-p)}.  
\end{equation}

\section{Derivation of the Stokes switching} \label{sec:switch}

In order to derive \eqref{eq:Aexp}, we substitute 
\begin{equation}
  R_N \sim \mathcal{S}(z) B(z) \e^{-\chi(z)/\eps},
\end{equation}
into the equation for the remainder \eqref{eq:Rneq}, where $B$ is given in \eqref{eq:harmB} and $\chi$ is given in \eqref{eq:u}. The pre-factor $\mathcal{S}$ is the switching function, which is expected to rapidly vary across Stokes lines in the limit $\ep \to 0$. Using the late terms \eqref{late}, this gives the expression for the switching function as
\begin{equation} 
  2\i \phi' \mathcal{S}' \e^{-\chi/\eps} \sim -\eps^N \left[ \frac{(-\chi')^2 \Gamma(N + \gamma + 1)}{\chi^{N+\gamma+1}} \right].
\end{equation}
Writing $\chi' = 2\i \phi'$ and converting $\mathcal{S}' = \chi' \dd{\mathcal{S}}{\chi}$, we have
\begin{equation} \label{eq:dSdchi}
  \dd{\mathcal{S}}{\chi} \sim - \frac{\eps^N \e^{\chi/\eps}\Gamma(N + \gamma + 1)}{\chi^{N+\gamma+1}}.
\end{equation}
By expanding $\Gamma(N + \gamma + 1)$ in the limit $N \to \infty$, it can be shown that the right hand-side of \eqref{eq:dSdchi} is algebraically small except if $N$ is chosen optimally, i.e. at the point where adjacent terms of the expansion are equal,
\begin{equation}
  \left\lvert \frac{\ep^{N+1}}{\ep^N} \frac{A_{N+1}}{A_N} \right\rvert \sim \left\lvert\epsilon \frac{N}{\chi} \right\rvert\sim 1,
\end{equation}
or $N \sim |\chi|/\ep$. At optimal truncation, it can be shown that there exists a boundary layer near Stokes lines specified by \eqref{eq:stokesline} where $\mathcal{S}$ incurs a jump of magnitude $2\pi \i/\ep^\gamma$. Returning to \eqref{eq:dSdchi}, we note this is identical to eqn (4.4) of Chapman \& Vanden-Broeck~\cite{chapman_2006_exponential_asymptotics} with a negated right-hand side and where their $N$ is our $N-1$. The derivation of the jump conditions is identical otherwise and yields \eqref{eq:Aexp}. More details of the optimal truncation and Stokes switching procedure can be found in \cite{chapman_1998_exponential_asymptotics}. 

\end{appendix}


\end{document}